\documentclass{aastex631}

\usepackage{graphicx}
\usepackage{subfigure}

\usepackage{graphicx}
\usepackage{float} 
\usepackage{enumitem}

\begin{document}
\title{Spectro-polarimetry of GRB 180427A: evidence for distinct emission sites with varying polarisation}

\author[0009-0009-2018-9457]{Rushikesh Sonawane}
\affiliation{Centre for High Performance Computing, Indian Institute of Science Education and Research Thiruvananthapuram, Thiruvananthapuram, 695551, India}
\affiliation{School of Physics, Indian Institute of Science Education and Research Thiruvananthapuram, Thiruvananthapuram, 695551, India}
\email{rushikesh23@iisertvm.ac.in}

\author{Shabnam Iyyani}
\affiliation{School of Physics, Indian Institute of Science Education and Research Thiruvananthapuram, Thiruvananthapuram, 695551, India}
\affiliation{Centre for High Performance Computing, Indian Institute of Science Education and Research Thiruvananthapuram, Thiruvananthapuram, 695551, India}
\email{shabnam@iisertvm.ac.in}

\author{Soumya Gupta}
\affiliation{Homi Bhabha National Institute, Anushakti Nagar, Mumbai Maharashtra-400094, India}

\author{Tanmoy Chattopadhyay}
\affiliation{Kavli Institute of Particle Astrophysics and Cosmology, Stanford University
452 Lomita Mall, Stanford, CA 94305, USA}


\author{Dipankar Bhattacharya}
\affiliation{ Department of Physics, Ashoka University, Sonepat, Haryana 131029, India}

\author{ Varun. B. Bhalerao}
\affiliation{Indian Institute of Technology Bombay, Mumbai, India}

\author{Santosh V. Vadawale}
\affiliation{Physical Research Laboratory, Ahmedabad, Gujarat 380009, India}

\author{G. C. Dewangan}
\affiliation{Inter-University Center for Astronomy and Astrophysics, Pune, Maharashtra 411007, India }







\begin{abstract}
The dynamics of the origin of gamma-ray emissions in gamma-ray bursts (GRBs) remains an enigma. Through a joint analysis of 
GRB 180427A, observed by the Fermi Gamma-ray Space Telescope and AstroSat's Cadmium Zinc Telluride Imager, we identify emissions from two distinct regions with varying polarisation 
properties. Time-resolved polarisation analysis reveals a synchronous evolution of the polarisation angle (PA) and fraction (PF) with two emission pulses, peaking with a delay of $ 5.09 \pm 0.29\, \mathrm{s}$. 
Spectral analysis indicates that the first pulse is characterised by a stronger blackbody component, while the second pulse exhibits a more prominent non-thermal spectrum (power law with an exponential cutoff).
Using a bottom-to-top approach through simulations, we decouple the polarisation properties of the individual spectral components, revealing polarisation fractions 
of 25\% - 40\% for the blackbody spectrum and 30\% - 60\% for the non-thermal spectrum. At a redshift of $z \sim 0.22$, the blackbody emission originates from the jet photosphere at $\sim$ a few $10^{11}\, \mathrm{cm}$, whereas the non-thermal emission arises 
from  an optically thin region at a few $10^{13}\, \mathrm{cm}$. The changing dominance of these emissions explains the observed PA shift of $60^\circ 
\pm 22^\circ$. The spectral cutoff at 1 MeV suggests pair opacity due to the jet's relatively lower bulk Lorentz factor ($\Gamma \sim$ a few tens). The high 
polarisation fraction and hard low energy spectral slopes ($\alpha > -0.5$) imply a top-hat jet structure observed off-axis, near the jet's edge. This off-axis viewing introduces anisotropy in the observed radiation within the viewing cone ($1/\Gamma$), accounting for the observed polarisation.

\end{abstract}





\section{Introduction} \label{sec:intro}

Gamma-ray Bursts (GRBs) are among the most energetic transient phenomena in the universe. These bursts have been extensively studied by various space-based gamma-ray observatories, including the {\it Neil Gehrels Swift} observatory \citep{gehrels2004swift}, the {\it Fermi} Gamma-ray Space Telescope \citep{meegan2009fermi}, and AstroSat's Cadmium Zinc Telluride Imager (CZTI, \citealt{Rao_etal_2017_Astrosat}). Understanding the nature of the prompt gamma-
ray emission in GRBs through spectroscopy remains a significant challenge, often hindered by degeneracies where multiple spectral models fit the same data. Polarisation observations of GRB prompt emission are pivotal in breaking these 
degeneracies, as they provide insights into the radiation mechanisms, jet structure, and viewing geometry \citep{Toma_etal_2009,Gill_etal_2021_review,Gill_etal_2021_pol_evol}. Notably, 
polarimetric observations of GRBs have been widely reported by {\it AstroSat} since 2015 \citep{vadawale2015hard} and POLAR between 2016–2018 \citep{POLAR_2018}. The integration of spectral data from {\it Fermi} with 
polarimetric measurements from AstroSat CZTI and POLAR are instruments in advancing our understanding of the GRB and its physics.

 It has been observed that GRBs can exhibit intrinsic polarisation. Yet, time-averaged measurements may appear unpolarised due to a varying polarisation angle \citep{2019ApJ_sharma,Burgess_etal_2019_pol,Kole_etal_2020_pol_catalog,Pol_catalog_2022,Gupta_etal_2024}. Additionally, off-axis observations, especially with 
 emissions consistent with photospheric, Compton-drag, or synchrotron emissions from random magnetic fields, can exhibit polarised signatures \citep{Chand2018,Sharma_etal_2020}. In this work, we present GRB 180427A as a case study demonstrating polarisation evolving synchronously with two emission pulses within a broadly fast-rise and exponential-decay (FRED)  like light curve.

On April 27, 2018, the Gamma-ray Burst Monitor (GBM) aboard the {\it Fermi} Gamma-ray Space Telescope triggered at 10:37:03.04 UT (Fermi MET 546518228), detecting GRB 180427A at (R.A., decl.) = (283.33$^{\circ}$, 70.30$^{\circ}$) with a $1^{\circ}$ uncertainty.
The burst was also localized to $R.A. = 271.01^{\circ}$, $decl. = 68.45^{\circ}$ using IPN triangulation \citep{2018GCN.22679....1H}.
The GBM light curve revealed a FRED-like pulse with a $T90$ duration of 26 seconds in the 50–300 keV range \citep{Fermi_GBM_180427A}, and a fluence of $4.92 \pm 0.06 \times 10^{-5}$ $\rm erg\,cm^{-2}$. Konus-Wind \citep{Konus_wind_180427A} and AstroSat CZTI \citep{Astrosat_180427A} also detected the burst. However, {\it Swift} Burst Alert Telescope (BAT) did not detect this burst as it was outside BAT's coded field of view \citep{IPN_180427A}. AstroSat CZTI measured a T90 = 22.4 s (40–200 keV) and recorded 981 Compton events (100–600 keV). No afterglow or redshift was reported for this burst.

The sections \ref{sec:obs_lc}, \ref{spec},
and \ref{sec:pol} detail the light curve, spectral, 
and polarisation analyses, respectively, followed by a discussion and summary in section \ref{sec:discussion} and section \ref{sec:summary} respectively.

\section{Lightcurve Analysis} \label{sec:obs_lc}

\begin{figure}[!b]
  \centering
  \includegraphics[width=1.0\textwidth]{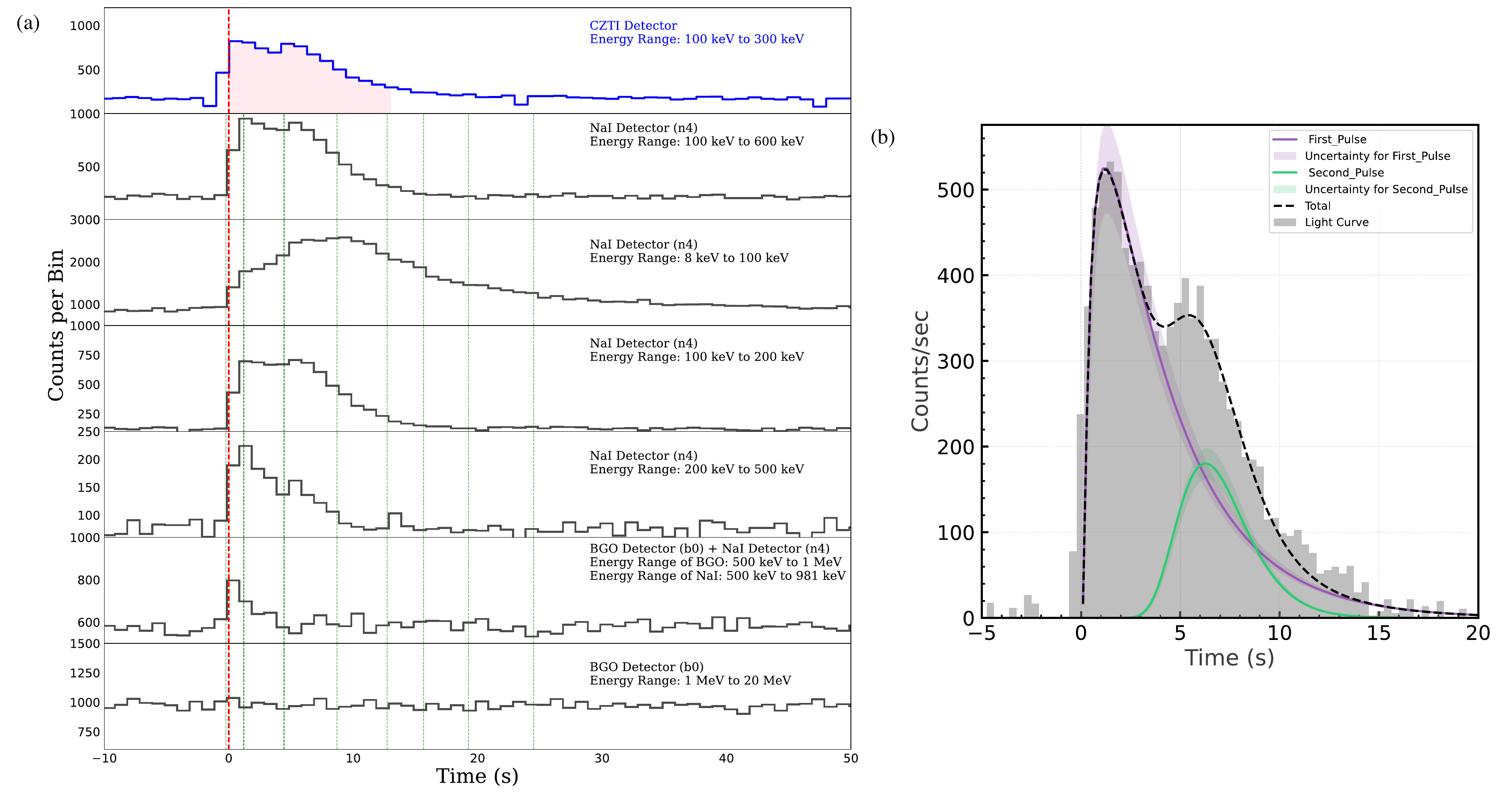}
  \caption{(a) The composite light curve of GRB 180427A is shown, with the top panel displaying the 100–300 keV single-event light curve from AstroSat CZTI. The pink-shaded region indicates the interval used for time-integrated polarisation analysis. The following panels show light curves from the {\it Fermi} NaI 4 and BGO 0 detectors, covering increasing energy ranges. The dashed red and green lines mark the {\it Fermi} GBM trigger time and the intervals for time-resolved spectral analyses, respectively. (b) The two peaks in the 100–600 keV background subtracted {\it Fermi}
  light curve are modeled by Norris pulses (magenta and green curves), with shaded regions representing $68\%$ confidence intervals. The gray histogram in the background shows the 0.37-second binned light curve for the same energy range.} 
  \label{lc_cross}
\end{figure}

Among the 12 sodium iodide (NaI) detectors and 2 bismuth germanate (BGO) detectors on the \textit{Fermi} Gamma-ray Burst Monitor (GBM), only NaI4 (n4) was positioned at an angle less than 
$60^\circ$, while BGO0 (b0) was the brightest BGO detector. These were utilised for the analysis. The  burst 
exhibited a broad pulse with two peaks and fast-rise exponential decay (FRED) light curve, as 
shown in Figure \ref{lc_cross}a. The composite light 
curve spans a broad energy range from 8 keV to 20 MeV. The top panel displays the single-event light curve from AstroSat CZTI in the 100-300 keV range, with the shaded pink region indicating the time interval used 
for time-integrated polarisation analysis. AstroSat CZTI allows to measure linear polarisation in the energy range 100 - 600 
keV using the technique of Compton polarimetry. Both AstroSat CZTI (100-300 
keV) and \textit{Fermi} GBM (100-600 keV: the energy band similar to CZTI polarisation analysis) light curves shown in the top 
two panels respectively in Figure \ref{lc_cross}a exhibit two distinct peaks. No significant emission was observed above 1 MeV in the BGO0 detector, suggesting a spectral cut-off or a steeper power law at higher energies.

The \textit{Fermi} background subtracted light curve binned in $0.37 \rm s$ in the 100-600 keV range was analysed using a model comprising two Norris emission pulses \citep{Norris_etal_2005}. A Norris emission pulse is given by:
\begin{equation}
    I(t) = K \exp\left(-\frac{\tau_1}{t} - \frac{t}{\tau_2}\right)
\end{equation}

where $K$ is the normalisation constant, $\tau_1$ is the rise time, and $\tau_2$ is the decay time after the pulse maximum. The model was fitted to the light curve using the Maximum Likelihood Estimation (MLE) method 
in the 3ML software \citep{Vianello_etal_2015}, yielding a reduced chi-square of 0.72. The two emission pulses are shown in Figure 
\ref{lc_cross}b as magenta and green curves. The total model, formed by combining the two pulses, is shown by the dashed black curve, which closely matches the two observed peaks in the light curve (100–600 keV), 
depicted in gray in the background of Figure \ref{lc_cross}b. The shaded regions around the curves represent the 68\% uncertainty level of model derived from the fit. The estimated parameter values for two emission pulses are given in table \ref{tab:Norris}.
We specifically analyze the 100–600 keV energy range as this matches the energy range in which the AstroSat CZTI can effectively measure polarisation.

The time delay between the obtained Norris pulses was determined by identifying the peak of each pulse and calculating the difference between them. 
    To account for uncertainties, multivariate sampling incorporating the covariance matrix was performed, simulating nearly $10^5$ instances of the Norris pulses. The averages of the distributions of the peak times of the first and second emission pulses are $t_{peak}^{pulse1} = 1.2$ s and $t_{peak}^{pulse2} = 6.3$ s respectively. 
    The time delay was estimated for each simulation. The mean and standard deviation of the time delay distribution are $t_{\text{delay}} = 5.09 \pm 0.29$s. 
    Further examination of the spectral composition and origin of these emission pulses is presented in the next section \ref{spec}.

\begin{table}[h]
    \centering
    \begin{tabular}{|c|c|c|}
    \hline
    Parameters & Pulse 1 & Pulse 2 \\
    \hline
    K & $(1.01^{+0.13}_{-0.12} ) \times 10^{3}$ & $(1.5^{+4.0}_{-1.1} ) \times 10^8$ \\
    $\tau_1$ & $0.41 \pm 0.09$ & $43 \pm 4$\\
    $\tau_2$ & $3.5 \pm 0.4$ & $0.92 \pm 0.1$ \\
    \hline
    \end{tabular}
    \caption{The parameter values estimated for the fitted Norris pulses over the light curve in the energy range 100 keV - 600 keV are listed above.}
    \label{tab:Norris}
\end{table}

\section{Spectral Analysis}
\label{spec}

The time-integrated and time-resolved spectroscopy of the burst were performed using \textit{Fermi} GBM data 
in the Multi-Mission Maximum Likelihood (3ML) software. The spectral parameters were estimated via 
the maximum likelihood estimation method, with the Akaike Information Criterion (AIC, \citealt{akaike1974}) applied to select the best-fitting model.

For time-integrated spectral analysis, the time interval identified to encompass the full burst emission is [-2, 45] seconds. The burst was best modeled using the Band function 
\citep{1993ApJ...413..281B} alone, which provided the lowest AIC value. The resulting Band function parameters are: low-energy power-law index $\alpha = 
-0.54 \pm 0.05$, peak energy $E_{peak} = 110 \pm 3 \, \rm keV$, and high-energy power-law index $\beta = 
-2.61 \pm 0.07$. The total burst fluence was found to be $6.25 \times 10^{-5} \, \rm erg/cm^2$ and $5.4 \times 10^{-5} \, \rm erg/cm^2$  in the observed energy range of 8 keV to 30 MeV and 10 keV to 1 MeV respectively. 

\begin{figure}[H]
  \centering
  \includegraphics[width=1.0\textwidth]{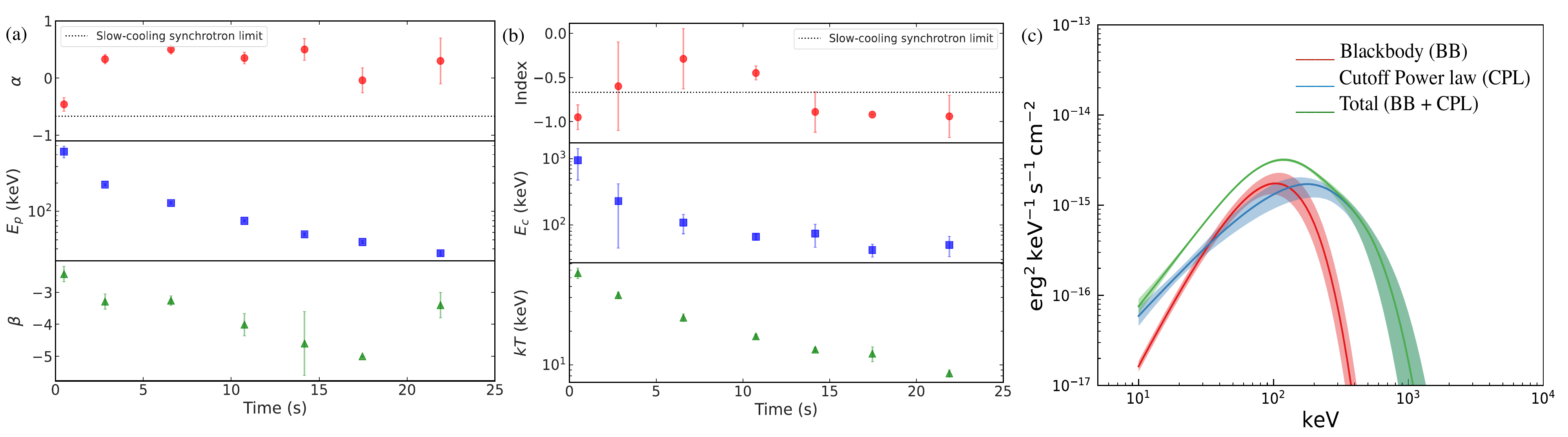}
  \caption{The temporal evolution of spectral parameters for (a) the Band function and (b) the Blackbody + Cutoff-powerlaw (BB + CPL) models is presented. For the Band function, the parameters $\alpha$, 
  $E_p$, and $\beta$ are shown in the upper, middle, and lower panels using red circles, blue squares, and green triangles, respectively. For the BB + CPL model, the power-law index, cutoff energy, $E_c$, and blackbody temperature, $kT$ are displayed similarly in the upper, middle, and lower panels. The dotted line in 
  the $\alpha$ and Index plots indicates the expected low-energy power-law index ($-0.67$) for slow-cooled synchrotron emission. (c) The $\nu F_{\nu}$ plot of the BB + CPL model obtained for one of the brightest bins (1.2 - 4.4 s) is shown. 
}   
  \label{fig:spec}
\end{figure}

For time-resolved spectroscopy, the light curve from the brightest NaI detector (n4) was binned using the Bayesian Block binning method, with a false alarm probability of $p_0 = 0.01$ \citep{scargle2013studies}. This method optimally segments the data into blocks where no significant variation from a constant 
rate is observed. As Bayesian blocks do not ensure a sufficient signal-to-noise ratio in each bin, we combined several bins during the main burst emission and excluded bins beyond 25 
seconds due to their very low signal-to-noise ratio. This process resulted in seven time intervals with adequate signal-to-noise 
ratios for constraining the spectral parameters, as illustrated in Figure \ref{lc_cross}a. The Band function alone provided the best fit for all intervals. The temporal evolution of spectral parameters is depicted in Figure \ref{fig:spec}a, where the photon index $\alpha$ is steeper than the "line of death" for synchrotron emission ($\alpha = -0.67$, \citealt{preece1998synchrotron}), suggesting a potential thermal component.

The light curve showed two distinct emission pulses, indicating the presence of two emission sites and thereby possibly two spectral components. To explore this further, we analysed the spectrum using two-component models: blackbody (BB) + Band and BB + cutoff power-law (CPL). Both models resulted in random residuals with $\Delta AIC \leq 5$ suggesting they are equally plausible models. Following Occam's Razor rule, the BB + CPL model was preferred due to fewer spectral parameters and the steep $\beta$ of the Band function, which aligns with the absence of emission above 1 MeV, as observed in the light curve. The temporal evolution of the spectral parameters, including power-law index, spectral cutoff energy ($E_c$), and blackbody temperature ($kT$), is shown in Figure \ref{fig:spec}b.
The $\nu F_{\nu}$ plot for the BB + CPL model, derived from one of the brightest time bins, is shown in Figure \ref{fig:spec}c. The shaded region represents the $68\%$ uncertainty level of the spectral model shape derived from the fit.

\subsection{Comparison between photon flux and count lightcurves} \label{Photon_Flux_LC}

The polarisation analysis is limited to 100–600 keV energy range and the burst light curve in this range reveals two distinct emission pulses, as discussed 
in Section \ref{sec:obs_lc} and shown in Figure \ref{lc_cross}b. To investigate the origin and spectral 
composition of these pulses, the photon flux light curves derived from the spectral analysis are compared with the corresponding count-rate light curves.

The spectral analysis indicates that the emission consists of two distinct spectral components, each peaking at different energies, as illustrated in Figure \ref{fig:spec}c. For instance, in the $\nu F_{\nu}$ plot of the brightest interval, the 
blackbody (BB) component peaks at $\sim 100$ keV, while the cutoff power-law (CPL) component becomes dominant at higher energies around $\sim 200$ keV. Notably, these 
peaks are well constrained with no significant overlap in their error regions. Furthermore, through out the burst duration CPL peaks at energies higher than the BB peak (Figure \ref{fig:spec}b). This confirms that the relative dominance of each component is strongly energy-dependent.

\begin{figure}[H]
  \centering
  \includegraphics[width=1.0\textwidth]{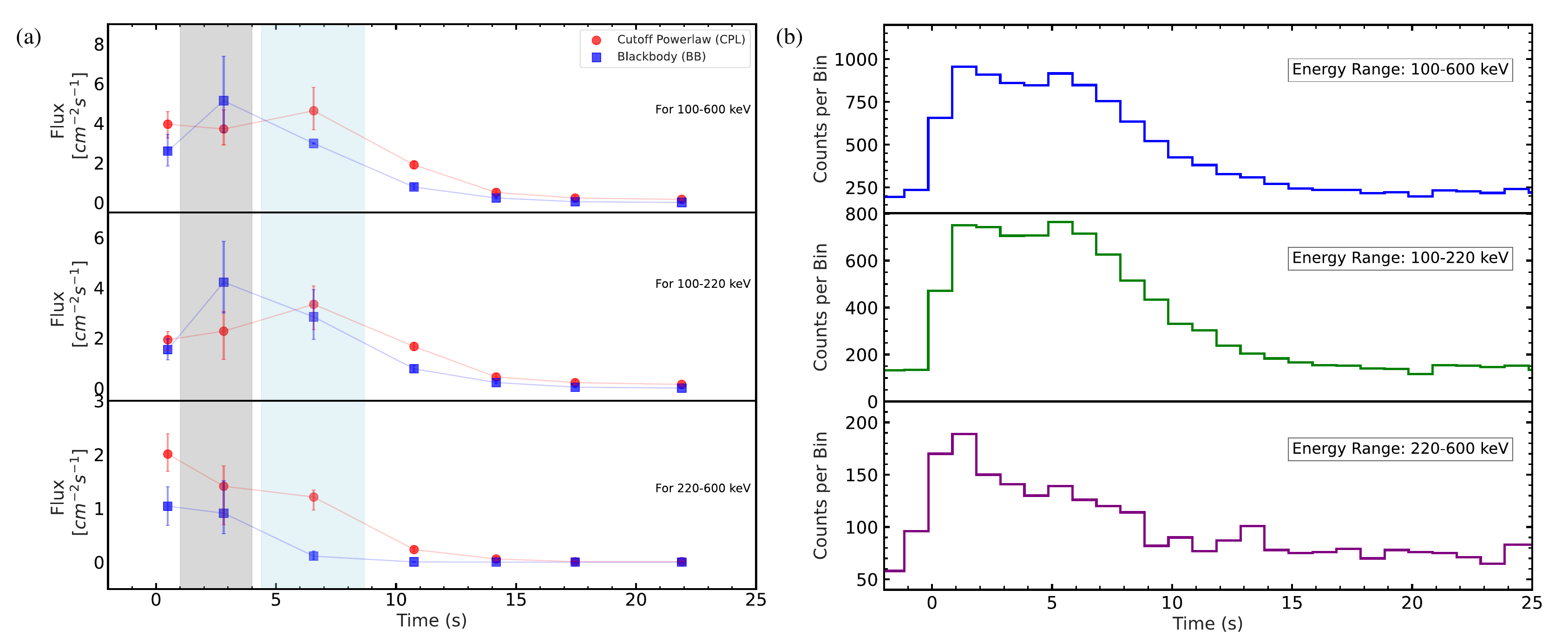}
  \caption{ (a) The temporal evolution of photon flux for the BB (blue squares) and CPL (red circles) components is shown across three energy ranges—100–600 keV, 100–220 keV, and 220–600 keV—in the upper, middle, and lower panels, respectively. The time intervals identified for BB and CPL dominance are shown in shaded grey and blue respectively. (b) Light Curves in the different energy ranges: 100–600 keV, 100–220 keV, and 220–600 keV—are in the upper, middle, and lower panels, respectively.
}   
  \label{fig:photon}
\end{figure}

Furthermore, throughout the burst duration, all time-resolved spectra consistently show the presence of both components (Figure \ref{fig:spec}b). Therefore, the two 
emission pulses observed in the light curve cannot be attributed exclusively to one component each. To investigate the spectral makeup of each pulse, we 
analysed the photon flux light curves and compared them with the corresponding count-rate light curves. This analysis is presented in the Figure \ref{fig:photon} where the photon flux light curves are shown on the left hand while the counts light curves in the same energy ranges are shown on the right hand side. It is now 
evident that in the 100–600 keV range, the BB component peaks at $t = 2.8$ s in the interval $ 1- 4$ s and the CPL component peaks at $t = 6.6$ s in the interval $4.4 - 8.7$ s (uppermost panel of Figure \ref{fig:photon}a) corresponding to the observed two peaks in the counts light curve (uppermost panel of Figure\ref{fig:photon}b). The identified time intervals, i.e the respective Bayesian block intervals, are highlighted as the shaded region in Figure \ref{fig:photon}a.  

Considering the overall spectral evolution showing a decrease in peak 
energy with time (Figure \ref{fig:spec}b), we subdivided the light curves in 100–600 keV range into two ranges: 100–220 keV and 220–600 keV. 
The division at 220 keV was conservatively chosen based on the temporal evolution of $kT$, where the BB peak (approximately $2.8 \, kT$ in the $F_{\nu}$ 
space) reaches a maximum of $\sim 200$ keV. This choice was made to provide a conservative threshold, considering both physically relevant quantities - 
photon flux and energy flux - where the division based on the $F_{\nu}$ space ensures that the main thermal energy contribution is adequately captured.
Photon flux light curves for the resolved 
energy ranges (middle and lower panels of Figure \ref{fig:photon}a) were generated and compared with the corresponding count-rate light curves (middle and lower 
panels of Figure \ref{fig:photon}b). The similarity in peak structures and overall behavior confirms that the features observed in the count light curves reflect the underlying photon flux variations of BB and CPL components.  

This approach enabled to identify the time intervals and energy bands where each component dominates, which is crucial for polarisation studies. Minimising the contribution of one component in a given interval helps avoid cancellation of polarisation signals that may arise when the components have different polarisation angles. Based on this analysis, the following observations are made.:

\begin{itemize}[itemsep=5pt, topsep=3pt, parsep=3pt, partopsep=3pt]
\item Each of the two emission pulses includes contributions from both the BB and CPL components.
\item The first pulse, identified in the interval $t = 1$–$4$ s, is primarily characterised by a stronger BB component. This prominence is more evident in the 100–220 keV range, where higher-energy CPL contributions are suppressed. The photon flux ratio $\frac{F_{ph,BB}}{F_{ph,CPL}}$ in 100–600 keV is $1.4$, and in 100–220 keV it is $1.8$.
\item The second pulse, occurring in the interval $t = 4.4$–$8.7$ s, is dominated by the CPL component. This is even more evident in the 220–600 keV range, where lower-energy BB contributions are minimised. The photon flux ratio $\frac{F_{ph,CPL}}{F_{ph,BB}}$ in 100–600 keV is $1.6$, and in 220–600 keV it is $10.5$.
\end{itemize}

In conclusion, the spectral analysis identifies two distinct components—BB and CPL—peaking at different energies. By constructing the respective photon flux light curves in 
100 - 600 keV energy band and comparing them with the observed count light curves, the spectral composition of each emission pulse is determined. The first pulse is 
BB-dominated, while the second is CPL-dominated\footnote{Hereafter, {\it “BB-dominated” } and {\it “CPL-dominated”} refer to where the BB or CPL component contributes relatively stronger in photon flux, even if the ratio is modest. This usage is intended for convenient reference to the relative contributions.}. Furthermore, we also identify the specific time intervals and energy bands 
where each component’s dominance is maximal, which is particularly useful for optimising the time and energy resolved polarisation analyses discussed in Section \ref{sec:pol}.



\section{Polarisation analysis} \label{sec:pol} 
\subsection{Methodology}
AstroSat CZTI is equipped to conduct polarisation studies in the energy range of 100–600 keV using the Compton polarimetry technique.
When polarised photons are incident on the detector, they tend to scatter perpendicularly to their electric field vector. A Compton polarimeter analyses the 
distribution of these scattered photons with respect to the azimuthal angles in the detector plane. This distribution is modeled using a cosine function of the form:
\begin{equation}
C(\phi) = A \cos\left[2\left(\phi - \phi_0 + \frac{\pi}{2}\right)\right] + B
\label{eq:cosine_equation}
\end{equation}
where $\phi_0$ is the polarisation angle (PA) of the incident photons as measured in the CZTI detector plane, A/B is the modulation factor ($\mu$), and $\phi$ represents the azimuthal angle in the detector plane. The polarisation 
fraction (PF) is calculated by normalising the modulation ($\mu$) with $\mu_{100}$, which is the modulation factor for $100\%$ polarised emission incident on the detector from the same direction as the 
GRB, with the same spectrum and detected Sky PA. The detector plane PA is converted to the sky PA in the sky reference frame, taking into account the satellite's 
orientation. The detector PA is the measured polarisation angle in the instrument’s local frame and Sky PA is the true polarisation angle in the celestial reference frame, independent of the detector, and represents the source’s intrinsic polarisation orientation. 

The methodology for polarisation measurement presented in this work follows the detailed description in \citealt{Pol_catalog_2022}. The Compton events for the burst is selected based on Compton kinematics and 
adjacent pixel criteria. The azimuthal distribution of Compton events is then background-subtracted using the combined distributions of Compton events from the pre- ($T_0 - 503$ to $T_0 -83$) and post-GRB ($T_0 + 97$ to $T_0 + 407$) 
regions. This background-subtracted distribution is then geometrically corrected by normalising it with the azimuthal distribution obtained from simulations when 
unpolarised GRB emission with the same spectrum and incident angle is incident on the detector. The final obtained azimuthal distribution of Compton events is modeled 
with a cosine function, as described in the equation \ref{eq:cosine_equation}, using Bayesian analysis employing Markov Chain Monte Carlo (MCMC) methods in PyMC3\footnote{\url{https://pypi.org/project/pymc3/}}. The posterior 
distributions for the polarisation fraction (PF) and polarisation angle (PA) were filtered to remove values where PF exceeded 100\%. The refined distributions were used to generate corner plots for PF and Sky PA. The 
reported PF and PA values correspond to the most probable values from these posterior distributions. Notably, the PA posterior distribution in this work showed a shoulder in addition to the primary peak. This 
feature is attributed to the large error bars in the azimuthal distribution (see Appendix \ref{double_hump} for further discussion). The reported PA value is derived from the most probable peak of the PA posterior distribution, modeled with a Gaussian function. The standard 
deviation of this fit is used as the 68\% credible region for PA, while the 68\% credible region for PF is directly taken from its posterior distribution. 

We adopt the null hypothesis as unpolarised radiation, represented by a  straight line model\footnote{A straight line model implies a uniform distribution of Compton events across all eight azimuthal scattering angles in the detector plane.}, and the alternate hypothesis as polarised radiation, modeled by a cosine function. The 
presence of polarisation, or the strength of evidence supporting the alternate hypothesis, is evaluated using the 
Bayes factor. To interpret the evidence grade in favour of the alternate hypothesis, we employ the Kass and Raftery's modified Jeffrey’s scale \citep{Kass_Raftery1995}. 

\subsection{Results of Polarisation Analysis}
\subsubsection{Time Integrated Polarisation Analysis}

\begin{figure}[b!]
  \centering
  \includegraphics[width=1.0\textwidth]{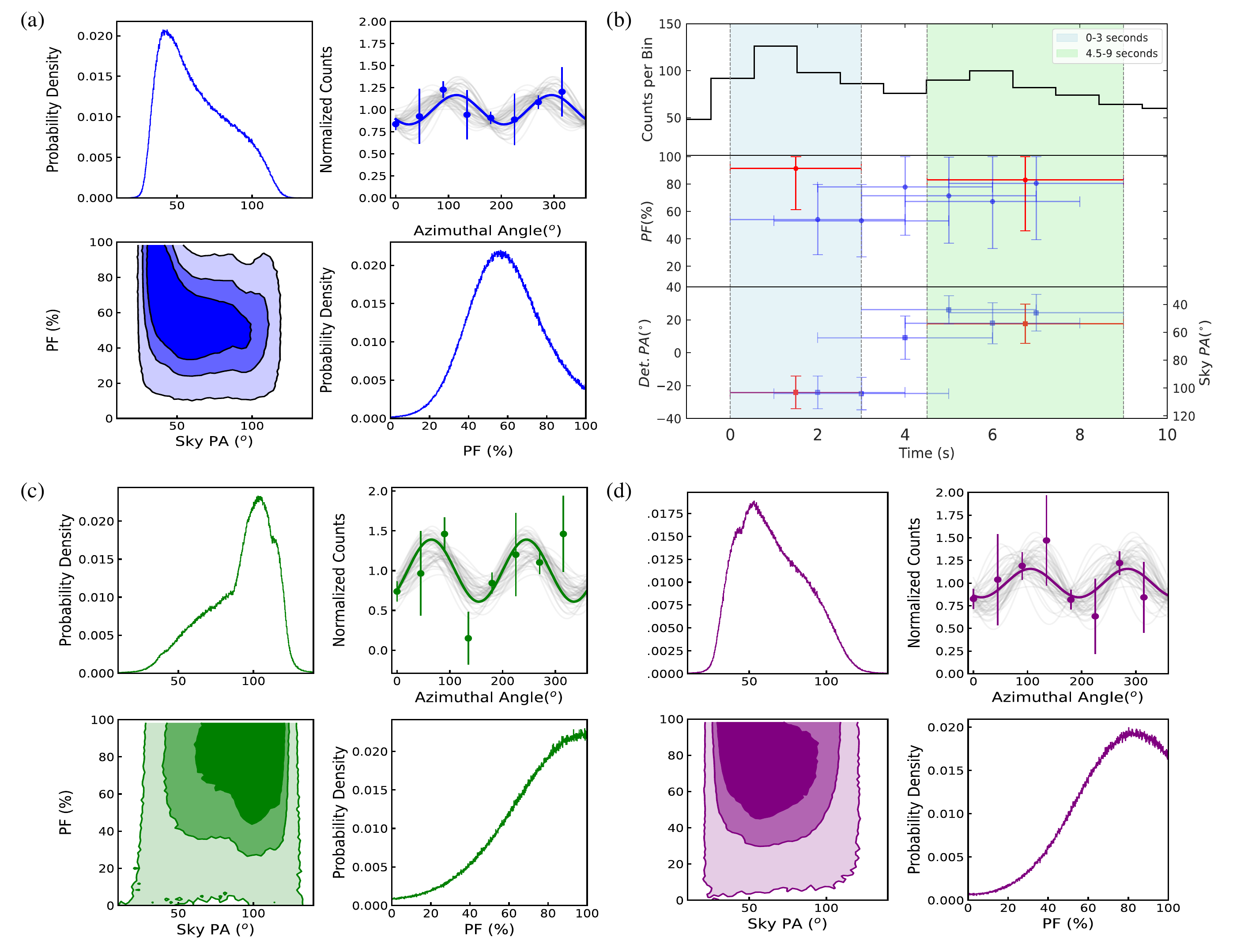}
  \caption{(a) The 2D histogram of PF and Sky PA are shown for time-integrated 
  polarisation analysis (0–13 s, 100–600 keV). (b) The top panel shows the Compton light curve from AstroSat CZTI in the 100–600 keV range. The middle and bottom panels 
  display the polarisation fraction (blue circles) and detector polarisation angle (blue squares) from a sliding window analysis with 4 s bins and 1 s steps. The second y-axis in the bottom panel represents the corresponding sky 
  polarisation angles. The time and energy resolved polarisation windows (0–3 s; 100–220 keV  and 4.5–9 s; 100–600 keV) are shaded in blue and green, with the 
  obtained PF and PA values shown in red circles and squares respectively. The 2D histograms of PF and Sky PA are shown for: (c) time and energy-resolved (0–3 s, 100–
  220 keV), and (d) time-resolved (4.5–9 s, 100–600 keV) in green, and purple corner plots, respectively. The contour lines in the 2D histograms represent 68\%, 95\%, and 99\% 
  credible regions for two parameters of interest. Each plot includes the most probable modulation curve (solid line) and 1000 randomly selected modulation curves  
  (shaded grey lines) from MCMC simulations in the top right. The obtained posterior distributions for PF and PA are shown in the bottom right and top left panels, respectively in each plot.}
  \label{Pol_results}
\end{figure}

Initially, we performed polarisation analysis on the time-integrated burst region, spanning 0 to 13 seconds, during which approximately 1020 Compton events were detected. The emission was found to be polarised, with a Bayes factor = 19.8 and $\Delta$WAIC = -15 \citep{watanabe2010asymptotic}, indicating positive evidence for polarisation. The polarisation fraction was measured as PF = $60 \pm 20\%$, and the sky PA was $42 \pm 10^\circ$ (Detector PA = $25 \pm 10^\circ$). The reported errors also account for the systematic errors mentioned in \cite{2019ApJ_Chattopadhyay} and \cite{2019ApJ_sharma}. Figure \ref{Pol_results}a presents the corner plot of the posterior distributions for PF and Sky PA, alongside the azimuthal distribution of Compton events and the most probable modulation curve.

\subsubsection{Time and Energy Resolved Polarisation Analysis}
\label{time_energy_pol}
To investigate the temporal evolution of polarisation properties, we performed a sliding window analysis with a 4-second window, incremented by 1 second. The 
resulting temporal variations in PF and PA are shown in the second and third panels of Figure \ref{Pol_results}b. The PA in the detector plane shifted from 
negative to positive values over time, with the PF constrained between 60\% and 80\% at a 68\% credible region for one parameter of interest.
The Bayes factor for each interval ranged between 2 and 3.2. Interestingly, the transition in PA from negative to positive occurred around $4.5$ seconds (lowest panel of Figure \ref{Pol_results}b). 
This timing is concurrent with the transition from the first to the second emission pulse, peaking at approximately $1.2$ s and $6.3$ s in the light curve, which are spectrally dominated by BB and CPL components respectively in 100 - 600 keV energy range. Notably, these timings emerge from independent light curve and polarisation analyses, reinforcing the physical connection between the spectral and polarisation evolution.
Hence, the presence of two distinct polarisation 
angles within the burst duration is likely linked to the two spectral components: BB and CPL.
To validate this interpretation with improved statistical significance, we performed a time- and 
energy-resolved polarisation analysis. This approach is essential, as it allows us to minimise the contribution of one component over the other in 
carefully selected time and energy intervals, thus avoiding possible cancellation of polarisation signatures due to differing PAs between components. 

To this end, we utilised the spectral analysis results along with a comparative study of the photon flux and count-rate light curves. 
Guided by the time and energy intervals identified in section \ref{Photon_Flux_LC}, polarisation measurements to isolate the intrinsic properties of 
each spectral component were attempted. However, during the second emission pulse, where CPL dominates, a limitation arose in 
the 220–600 keV band: despite the clear spectral dominance of the CPL component, the detected photon count was insufficient to produce a statistically significant number of Compton events for reliable polarisation analysis. As a result, we 
conducted the polarisation measurement for the second emission pulse in the full 100–600 keV band over the time interval 4.5–9 s—where the CPL component still 
maintains dominance, as detailed in Section \ref{Photon_Flux_LC}. Finally, based on the sliding window analysis, spectral results, and the requirement for adequate Compton statistics, we divided the Compton light curve into two intervals: 0–3s (100–220 keV) and 4.5–9s (100–600 keV), as shown in Figure~4(b). This segmentation 
enabled us to make the most statistically significant polarisation measurements to probe the PA evolution over time.
The analysis identified polarised emission in both intervals, with Bayes factor 5 and 10, and $\Delta$WAIC values of -5.49 and -12, 
respectively indicating a positive evidence for polarisation. The corner plots of the posterior distributions for PF and sky PA obtained for these temporal and energy-resolved regions, along with the azimuthal distributions of Compton events and the most probable modulation curves, are shown in Figures \ref{Pol_results}c and \ref{Pol_results}d. The most probable PA in the detector plane 
shifted from $-27 \pm 9^\circ$ to $+25 \pm 11^\circ$, aligning with the temporal evolution of the two emission pulses observed in the light curve. This confirms that both spectral components 
are polarised but with different polarisation angles.

\subsubsection{Decoupling of Polarisation of Spectral Components}
The time- and energy-resolved polarisation analysis presented in Section 4.2.2 yields a Bayes factor $> 3.2$ in each case, confirming the presence of polarised radiation with a shift in polarisation angle (PA). However, due to the continuous presence of both the BB and CPL components throughout the burst, their individual polarisation fractions (PFs) cannot be directly constrained. Firstly we summarise the key observational insights:
\begin{enumerate}
\item When the BB component dominates, the net PA in the detector plane is negative, and when the CPL component dominates, the net PA is positive.
\item As the dominance of the spectral components shifts over time, the PA in the detector plane transitions from negative to positive. 
\item The change in PA in the detector plane, $\Delta PA$, over time is $60 \pm 22^\circ$. 
\item The photon flux ratio of BB to CPL in the first interval (0–3 seconds) is $1.8$ in the 100–220 keV range.
\item In the CPL-dominated second interval (4.5–9 seconds), the photon flux ratio of CPL to BB becomes $2.2$ in the 100–600 keV range.

Note that these photon flux ratios are calculated for the final time intervals selected for the time- and energy-resolved polarisation analysis, as described in Section \ref{time_energy_pol}.
\end{enumerate}

To estimate the individual polarisation fractions of the components, we performed simulations following the 
setup outlined in Appendix \ref{double_hump} of this section, using the Geant4 simulation framework described therein. Using point (1), we fix the sky PAs 
of the BB and CPL components in the simulations to their most probable values derived from the 
polarisation analysis in the 0–4 s (100–220 keV) and 4.5–9 s (100–600 keV) intervals, respectively (Section 4.2.2), corresponding to negative and positive detector plane PAs. 

The simulation models polarised emission as a mix of BB and CPL components with fixed PAs and varying PFs. For each PF combination, we examine whether the 
resulting net polarisation—given the photon flux ratio of BB to CPL are $1.8$ (BB-dominated) and $0.45$ (CPL-
dominated) in the energy ranges 100 - 220 keV and 100 - 600 keV respectively —can reproduce the observed PA transition from negative to positive (point 2) including the observed change in polarisation angle, $\Delta PA$, in the detector plane (point 3) while maintaining a statistical significance with Bayes factor $> 3.2$.

We explore PF values from $20\%$ to $40\%$ for the BB component and $20\%$ to $70\%$ for the CPL component in increments of $5\%$. The upper limit for the search 
of BB PF is set at $40\%$, consistent with theoretical expectations from models such as 
\citealt{Lundman_etal_2014}, which suggest that BB emission can exhibit a maximum polarisation fraction 
of $\sim40\%$ \citep{Gill_etal_2021_review}. Considering PF values outside the above mentioned ranges, in the simulation, fail to reproduce the observed net PA behaviour described in point (1), and are therefore excluded.

PF combinations that reproduce all three observational constraints (points 1–3) in the given conditions of photon flux ratios (points 4-5) are considered viable. As a result, we constrain the PF of the BB component to lie within $25\%$–$40\%$, and the PF of the CPL component within $30\%$–$60\%$.

\section{Discussion} \label{sec:discussion}
In this section, we analyse the observational results, spanning spectral, temporal, and polarisation analyses, 
to construct a consistent physical interpretation of GRB 180427A. The discussion is organised into subsections that sequentially address the viewing geometry, estimate 
key outflow parameters and redshift constraints, and finally propose a plausible physical scenario explaining the burst’s observed features.
\subsection{Inference of Viewing Geometry}
\label{viewing_geometry}
\begin{figure}[H]
    \centering
     \includegraphics[width=1.0\textwidth]{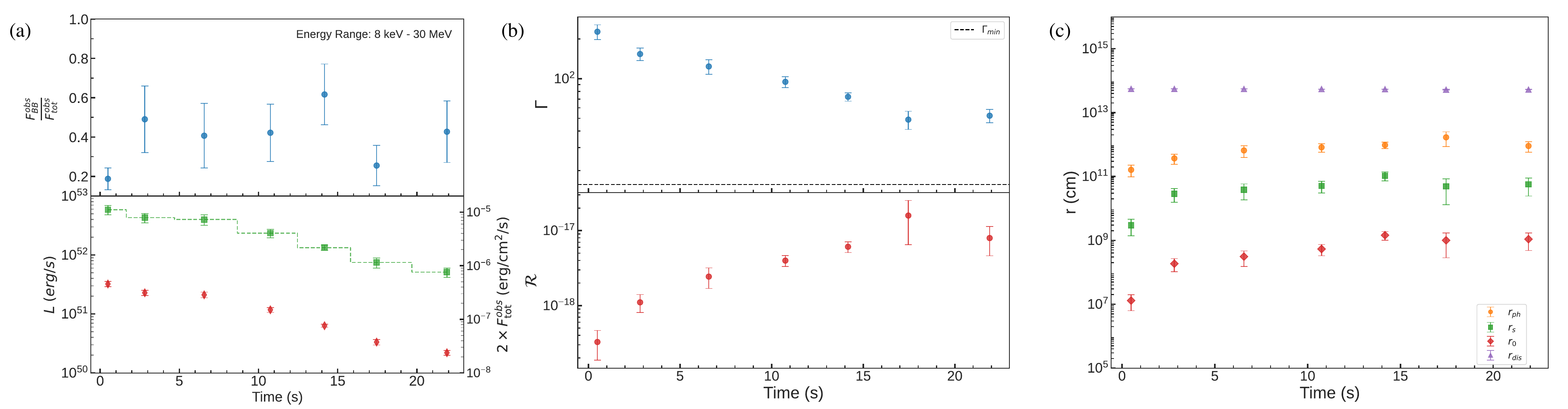}       
    \caption{The temporal evolution of the jet's outflow dynamics, based on the observed thermal component, is presented. (a) The ratio of observed blackbody flux to total flux, $F_{BB}^{obs}/F_{tot}^{obs}$, 
    is depicted in blue circles in the upper panel. In the lower panel, the burst luminosity $L$ (for a redshift $z = 0.05 $) is shown in red diamonds, while the total flux $2 \times F_{tot}^{obs}$ expected on-
    axis is plotted on the second y-axis in green square. (b)  The evolution of the bulk Lorentz factor ($\Gamma$) and the expected $\Gamma_{min}$ are plotted in blue circles and a dashed black line, 
    respectively, in the upper panel. The lower panel shows the evolution of ${\cal{R}}$ (red circles) over time.  (c) The temporal evolution of key radii—nozzle radius ($r_0$), saturation radius 
    ($r_s$), photospheric radius ($r_{ph}$), and dissipation radius ($r_d$)—are plotted using red diamonds, green squares, orange circles, and purple triangles, respectively. 
    }
    \label{fig:outflow}
\end{figure}

The presence of a prominent thermal component, contributing approximately observed blackbody flux $F_{BB}^{\mathrm{obs}} \sim 20\%$ to $60\%$ of the total observed flux, $F_{\mathrm{tot}}^{\mathrm{obs}}$, estimated in the energy range 8 keV - 30 MeV throughout the burst 
(Figure \ref{fig:outflow}a), along with the hard spectral slopes ($\alpha > -0.5$) (Figure \ref{fig:spec}a), points to photospheric emission and a 
baryon-dominated outflow \citep{Acuner_etal_2019, Acuner_etal_2020}. The non-thermal emission (CPL) can be attributed to optically thin radiation coming 
from a dissipation site well above the photosphere as indicated by the observed time delay, $t_{delay}$, between the two emission pulses revealed through light 
curve and spectral analyses where the first pulse is dominated by BB while the second pulse is dominated by CPL (refer section \ref{Photon_Flux_LC}). When viewed on-
axis, in baryon dominated outflows, blackbody and cutoff power-law components are expected to be unpolarised, regardless of whether the CPL arises from synchrotron emission in random magnetic fields or inverse Compton 
scattering. Therefore, the observed polarisation indicates a strong anisotropy of radiation within the observed view cone of 1/$\Gamma$ where $\Gamma$ is the bulk Lorentz factor \citep{Toma_etal_2009, Chand2018, Gill_etal_2021_review}.

In a structured jet model—where both the luminosity and bulk Lorentz factor decrease with increasing angle from the jet axis - an off-axis viewing geometry typically results in a softer lower energy spectral slopes in the observed overall spectrum due to reduced Doppler boosting and lower luminosity \citep{Lundman_etal_2013, Lundman_etal_2014}. However, this expectation is inconsistent with the observed hard low-energy spectral index, $\alpha > -0.5$ (Figure \ref{fig:spec}a).

For GRB 180427A, the combination of a hard $\alpha$ and significant polarisation, thus, points to the following plausible scenario: a top-hat jet structure viewed near its edge. Such a configuration can lead to significant anisotropy in the radiation field within the observer’s viewing cone ($\sim 1/\Gamma$), resulting in the observed polarisation in both the BB and CPL components, while retaining the spectral hardness ($\alpha$) characteristic of an on-axis view in a photosphere-dominated emission scenario \citep{Waxman2003, Toma_etal_2009, Chand2018}. 

We note, however, that more realistic jet structures, such as those considered in \citet{Lundman_etal_2014}, involve power-law declines 
in Lorentz factor beyond the jet core. In such cases, the combination of significant polarisation and hard 
spectral slope would favour scenarios where the jet’s angular structure decreases with relatively steep power-law indices outside the core.


\subsection{Estimation of Outflow Parameters and Redshift Constraint}
\label{outflow_params}
By adopting a top-hat jet model, neglecting high-latitude emission, and assuming a continuously active central engine during the burst, the thermal component allows us to estimate key outflow parameters such as the Lorentz factor 
$(\Gamma)$, the nozzle radius $(r_0)$, the saturation radius $(r_s)$, and the photospheric radius $(r_{ph})$, 
following the methodology described in \citet{pe2007new, iyyani2013variable}. Notably, the derivations in \citet{pe2007new} assume an on-axis viewing geometry. 
Furthermore, the calculations are performed under the condition that the photosphere lies well above the saturation radius—an assumption that holds in this case, 
where the observed spectrum is consistent with a blackbody plus a cutoff power-law component. The dominant pulse in the CPL component appears after a delay relative to the 
thermal peak, indicating that its emission region lies above the photosphere, where the blackbody component originates. The significant flux observed in the CPL 
component suggests that a substantial fraction of the burst's kinetic energy is available at the dissipation 
site, which is consistent with a scenario where the photosphere forms during the coasting phase—i.e. beyond the saturation radius—as predicted by the classical fireball model \citep{meszaros2006gamma,Iyyani2018}. 

From section~\ref{viewing_geometry}, we inferred that the top-hat jet is viewed near its edge. In this configuration, the observer sees emission from within an angle of 
$1/\Gamma$ centered on the viewing angle $\theta_v$, which coincides with the jet opening angle $\theta_j$. Since the 
line of sight lies at the jet boundary, only half of the relativistic beaming cone intersects the emitting region, resulting in an observed flux that is approximately half of 
what would be seen under on-axis viewing conditions. Therefore, to apply the equations from \citet{pe2007new}, 
the energy fluxes—both the blackbody and the total—are corrected by a factor of two to approximate their on-axis equivalents. The corresponding on-axis total flux, 
$2 \times F_{tot}^{obs}$, is shown in the lower panel of Figure~\ref{fig:outflow}a. These corrected values are then used to compute the burst luminosity, $L$, and to derive the outflow parameters.

The expressions for the outflow parameters are as follows:\\ 
(a) Lorentz factor $\Gamma$\\
\begin{equation}
    \Gamma \cong \left[(1+z)^2 d_L Y \frac{(2 \times F_{tot}^{obs}) \sigma_T}{2m_pc^3 {\cal{R}}}\right]^{1/4}
\label{Gamma}
\end{equation}
(b) Nozzle radius $r_0$\\
\begin{equation}
    r_0 \cong \frac{d_L}{(1+z)^2} \left(\frac{2\times F_{BB}^{obs}}{Y 2\times F_{tot}^{obs}}\right)^{3/2} {\cal{R}}
    \label{r0}
\end{equation}
(c) Saturation radius $r_s$
\begin{equation}
    r_s = \Gamma r_0
    \label{rs}
\end{equation}
(d) Photospheric radius $r_{ph}$
\begin{equation}
    r_{ph} = \frac{L\sigma_T}{8 \pi m_p \Gamma^3 c^3} = \frac{Y d_L^2 (2 \times F_{tot}^{obs}) \sigma_T}{m_p \Gamma^3 c^3}
    \label{rph}
\end{equation}
\noindent 
where $z$ is the redshift, $d_L$ is the luminosity distance, $Y$ is the inverse of radiation efficiency of the burst which is taken as $50\%$ \citep{racusin2009_jet, Racusin_etal_2011}, $\sigma_T$ is the Thompson cross-section and 
\begin{equation}
    {\cal{R}} \equiv \left(\frac{2\times F_{BB}^{obs}}{\sigma_{SB}T^4}\right)^{1/4}
    \label{R}
\end{equation}
\noindent
where $\sigma_{SB}$ is the Stefan-Boltzmann constant and $T$ is the observed temperature of BB component. 

The dissipation radius is estimated as
\begin{equation}
    r_{dis} =\frac{c t_{peak}^{pulse2}}{1 - \beta \cos\theta_v}
    \label{rdis}
\end{equation}
\noindent
where $c$ is the speed of light, $t_{peak}^{pulse2} = 6.3$ s is the peak time of the second emission pulse (or the delayed emission pulse), $\beta = v/c = \left(1 - 1/\Gamma^2\right)^{1/2}$ where $v$ is the velocity of the outflow and $\theta_v \equiv \theta_j$ 
when the line of sight is along the edge of the jet. An average jet opening angle of approximately $\sim 5^\circ$ 
is typically estimated for long GRBs detected by Fermi GBM \citep{Racusin_etal_2011}, and this value is used in the above expression to estimate $r_{dis}$.

To evaluate the outflow parameters, it is essential to constrain the possible redshift $z$ of the GRB. A lower limit on $z$ is estimated by considering the condition that 
the nozzle radius of the jet, throughout the burst, must remain above $10^6$ cm—the approximate event horizon radius of lower stellar-mass black hole with a mass $\sim 7\,M_{\odot}$, 
as typically assumed in collapsar scenarios \citep{Sharma_etal_2021}. According to Equation~\ref{r0}, for a given $z$, the nozzle radius $r_0$ scales with 
$(F_{BB}^{obs}/F_{tot}^{obs})^{3/2}$ and ${\cal R}$. As shown in the upper panel of Figure~\ref{fig:outflow}a, the ratio $(F_{BB}^{obs}/F_{tot}^{obs})$ remains relatively 
stable across the burst, while ${\cal R}$ increases steadily (lower panel of Figure~\ref{fig:outflow}b). To 
estimate the lower bound on $z$, we use the first time bin—where ${\cal R}$ is minimum — in the expression for $r_0$ (Equation \ref{r0}), setting $r_0 = 10^6$ cm. This yields a lower limit of $z_{low} = 0.03$. 

The upper limit on the redshift $z$ is estimated by requiring that the photospheric radius remains greater than 
the saturation radius throughout the duration of the burst, i.e., $r_{ph} > r_s$ \footnote{In the classical fireball model \citep{meszaros2006gamma,Iyyani2018}, thermal emission originates from the photosphere and non-thermal emission from dissipation of kinetic energy at larger radii. When the photosphere forms below the saturation radius ($r_{ph} < r_s$), most energy remains in photons, leading to thermal dominance and minimal kinetic energy, with no further acceleration beyond $r_{ph}$. In contrast, GRB 180427A shows comparable thermal and non-thermal components, implying the photosphere forms in the coasting phase ($r_{ph} > r_s$), where sufficient kinetic energy is available to power non-thermal emission, while thermal flux is reduced by adiabatic cooling $\propto (r_{ph}/r_s)^{-2/3}$ which eventually governs their relative contributions \citep{iyyani2013variable,iyyani2015extremely}.}. Substituting the expressions for the 
photospheric and saturation radii from Equations~\ref{rph} and \ref{rs} into this condition yields a constraint on the upper limit of the Lorentz factor.
\begin{equation}
    \Gamma < \left( \frac{d_L^2 Y 2 \times F_{tot}^{obs} \sigma_T}{2m_p r_0 c^3}\right)^{1/4}
\end{equation}
\noindent
Substituting for the expression of $\Gamma$ (equation \ref{Gamma}) in the above equation, leads to the relation that 
\begin{equation}
    \frac{d_L}{(1+z)^2} \cong \frac{r_0}{{\cal{R}}}
\end{equation}
The nozzle radius of the jet typically evolves to a maximum value comparable to the size of the core of a Wolf–Rayet 
progenitor star, which is approximately $10^9$ cm \citep{iyyani2013variable, Iyyani_etal_2016}. Since $d_L$ 
is inversely proportional to 
${\cal R}$, we adopt the minimum value of ${\cal R}$ for this estimate. Substituting these quantities yields an upper limit of $z_{upper} = 0.4$. Thus, the possible range of redshift for the burst is found to be $[0.03 - 0.4]$. 


Using the $\Lambda_{CDM}$ cosmology with $H_0 = 67.4 \pm 0.5 , \rm km , s^{-1} , Mpc^{-1}$, $\Omega_m = 0.315$, and $\Omega_{vac} = 0.685$ \citep{2020A&A...641A...6P}, the burst luminosity, $L$ (Figure~\ref{fig:outflow}a, lower panel), and the outflow parameters (Figures~\ref{fig:outflow}b and c) are estimated assuming an average redshift of $z = 0.22$, of the inferred possible range of $z$. The luminosity decreases from $3.5 \times 10^{51} \, \rm erg/s$ to $2 \times 10^{50} \, \rm erg/s$ over time with an average around $1.4 \times 10^{51}\, \rm erg/s$.
The $\Gamma$ decreases from $255$ to $40$ (Figure \ref{fig:outflow}b) with an average around $110$, consistent with recent findings by \citealt{Dereli_etal_2022} of lower values of $\Gamma$. The 
evolution of key outflow radii, such as the nozzle radius ($r_0$), saturation radius ($r_s$), and photospheric radius ($r_{ph}$), is shown in Figure \ref{fig:outflow}c. For $z = 0.22$, $r_{ph}$ averages around $8 \times 10^{11}$ cm while dissipation 
radius $r_{dis}$ is around $ 5.4 \times 10^{13} \, \rm cm$. The nozzle radius evolves from $6 \times 10^6$ cm to $2 \times 10^9$ cm with an average around $6.5 \times 
10^8$  cm while $r_s$ is around $4.7 \times 10^{10} \, \rm cm$.  These trends align with previous studies \citep{iyyani2013variable, Iyyani_etal_2016}. Furthermore, the high-energy cutoff observed in the CPL 
component at sub MeV range could be attributed to $\gamma \gamma$ pair production, indicative of a relatively low bulk Lorentz factor ($\Gamma$). In contrast, for 
$\Gamma$ values of several hundreds, such a cutoff would be generally expected at much higher energies, typically in the tens of MeV to GeV range 
\citep{Ackermann_etal_2012,vianello2018bright}. The burst light curve shows no significant emission above $1$ MeV (lowest panel of 
Figure \ref{lc_cross}a). Taking this as the observed maximum photon energy and using burst luminosity, dissipation radius for a redshift $z=0.22$, we estimate a lower limit on the Lorentz factor
in the optically thin region as $ \Gamma_{\text{min}} \sim 16$ \citep{lithwick2001lower} marked in the upper panel of Figure \ref{fig:outflow}b. 

Considering the upper and lower limits of the possible redshift, the maximum values of the outflow parameters 
obtained are: $\Gamma \sim 328$, $r_0 \sim 3 \times 10^9 \, \rm cm$, $r_s \sim 1 \times 10^{12} \, \rm cm$, $r_{ph} \sim 5 \times 10^{12} \, \rm cm$ and $r_{dis} \sim
5.6 \times 10^{13} \, \rm cm$ while the minimum values are: $\Gamma \sim 23$, $r_0 \sim 1 \times 10^6 \, \rm cm$, $r_s \sim 2 \times 10^7 \, \rm cm$, 
$r_{ph} \sim 2 \times 10^{9} \, \rm cm$ and $r_{dis} \sim 4 \times 10^{13} \, \rm cm$. The burst luminosity ranges from $3 \times 10^{48} \, \rm erg/s$ to $2 \times 10^{52} \, \rm erg/s$ over the possible range of redshift. 

\subsection{Physical Scenario}
Combining the spectro-polarimetric analysis results with the inferences from the previous sections, the following physical picture emerges for GRB180427A. The burst occurs 
at a redshift of $z \sim 0.22$ with a luminosity of $\sim 10^{51}$ erg/s. A relativistic baryon dominated outflow with $\Gamma \sim 110$, powered by a compact central engine formed via the 
core-collapse of the progenitor star, expands ballistically from an initial radius of $\sim 10^8$ cm. In this baryon-dominated outflow, the photosphere forms 
during the coasting phase at a radius of $\sim 10^{11}$ cm, while the jet's kinetic energy is dissipated at a larger radius of $\sim 10^{13}$ cm in the optically thin region.
The analysis indicates that the jet lacks significant angular structure and is consistent with a top-hat jet model. The observer's line of sight lies along the edge 
of the jet, such that only half of the $1/\Gamma$ cone is filled with radiation. This edge-on viewing geometry induces anisotropy in the observed radiation, resulting 
in measurable linear polarisation. Moreover, this geometry also explains the observed $\sim 5$ s delay between the thermal emission from the photosphere and the 
non-thermal cutoff power-law emission arising from the dissipation site.

\section{Summary}
\label{sec:summary}
In summary, the spectro-polarimetric analysis of GRB 180427A identifies two distinct emission components originating from different regions within the baryon-dominated outflow with a relatively low Lorentz factor, observed off-axis near the edge of the top-hat jet resulting in observable polarised radiation. The light curve and 
spectral analyses indicate a time delay between the observed peaks of the two emission pulses, with the first showing a stronger contribution from blackbody radiation from the photosphere (at $\sim$ a few $ 10^{11}$ cm) and the second showing a more significant contribution from CPL emission from the optically thin region (at a few $10^{13}$ cm) respectively. 
The off-axis viewing of the jet along its edge causes these emissions to appear as separate peaks in the light 
curve, with a delay of approximately $5$ seconds. The temporal variation in the dominance of these emissions results in changes to the observed polarisation angle with time, highlighting their distinct polarisation properties.

\section*{acknowledgments}
\begin{acknowledgments}
We express our gratitude to Prof. Felix Ryde and Prof. A. R. Rao for their enlightening discussions and invaluable insights. This publication utilizes data from the AstroSat mission, operated by the Indian Space Research Organisation (ISRO), and archived at the Indian Space Science Data Centre (ISSDC).The CZT-Imager was developed through a collaborative effort involving multiple institutes in India, including the Tata Institute of Fundamental Research (TIFR), Mumbai; the Vikram Sarabhai Space Centre, Thiruvananthapuram; the ISRO Satellite Centre (ISAC), Bengaluru; the Inter-University Centre for Astronomy and Astrophysics (IUCAA), Pune; the Physical Research Laboratory, Ahmedabad; and the Space Applications Centre, Ahmedabad. This research also incorporates data from the Fermi mission, accessed via the High Energy Astrophysics Science Archive Research Center (HEASARC) Online Service provided by NASA’s Goddard Space Flight Center. Additionally, the Geant4 simulations presented in this paper were conducted using the high-performance computing resources at the Inter-University Centre for Astronomy and Astrophysics (IUCAA). S.I. is supported by the DST INSPIRE Faculty Scheme (IFA19-PH245) and SERB SRG grant (SRG/2022/000211). 
\end{acknowledgments}

\appendix
\section{Double Humped Posterior of Polarisation Angle}
\label{double_hump}

AstroSat's mass model is simulated using the Geant4 toolkit \citep{GEANT4:2002zbu}, which includes detailed geometries for all the payload and the satellite bus \citep{2019ApJ_Chattopadhyay,2021JApA...42...93M}. The study of polarisation of off-axis sources are challenging, as the polarisation properties can be altered due to interactions with other payload components and the CZTI housing elements. By conducting a Geant4  simulation, we can trace each photon as it interacts with the detector, allowing for a more accurate analysis of its polarisation characteristics.

\begin{figure}[htbp]
\centering
 \includegraphics[width=1.0\textwidth]{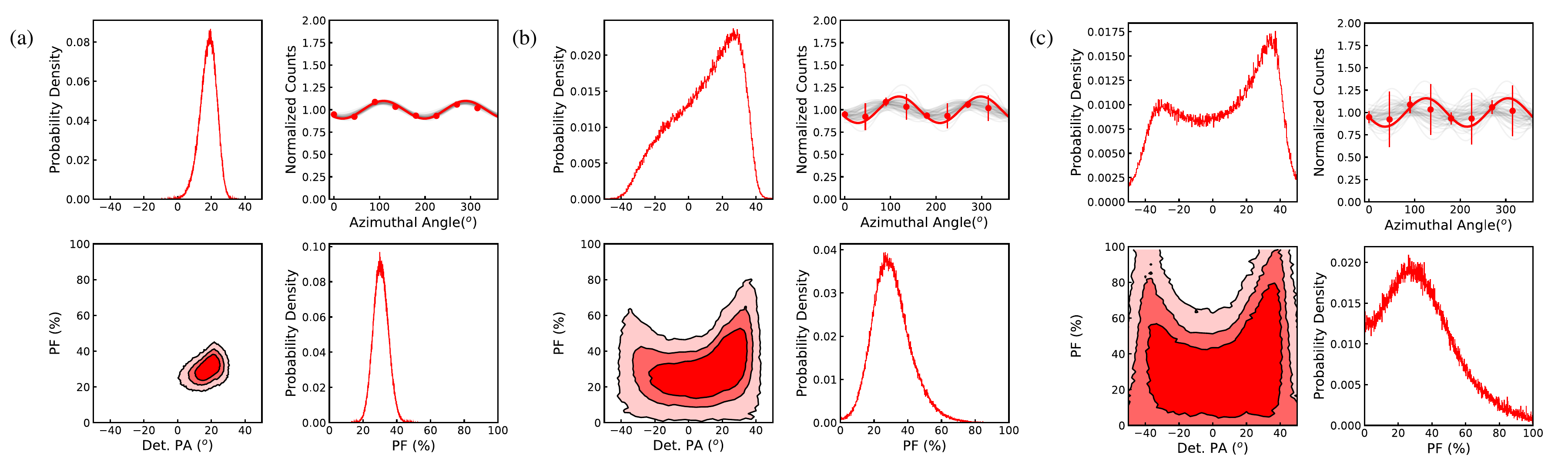}
    \caption{ The results of the polarisation analysis for various cases are presented, where the errors on the simulated modulation data points are (a) low, (b) moderately high, and (c) relatively large (comparable to the observations). For each case, the posterior distributions of the polarisation angle (PA) and polarisation fraction (PF), along with the corresponding 2D histograms with contour lines corresponding to 68\%, 95\% and 99\% credible regions for two parameters of interest are shown.} 
    \label{fig:simulation}
\end{figure}

Geant4 simulations were performed to investigate the double-humped posterior distributions of the polarisation angle observed in the data analysis of GRB 180427A. The simulations incorporated spectral information to direct photons from both the blackbody (BB) and cutoff power-law (CPL) components onto the detector.
In total, $8 \times 10^5$ photons were simulated to interact with the detector. For instance, we considered the scenario of BB-to-CPL photon flux ratio of $0.45$. The BB and CPL components were assigned polarisation fractions of $30\%$ and $60\%$, respectively, with polarisation angles in the sky plane of $112.4 \pm 2 ^{\circ}$ (detector PA = $-32 \pm 2 ^\circ$) and $38.8 \pm 2^\circ$ (detector PA = $28 \pm 2^\circ$).

Due to the large number of simulated photons, the number of Compton scattering events detected was approximately $10^4$. This resulted in a modulation curve with significantly reduced statistical errors (Figure \ref{fig:simulation} a). Analysis of this modulation curve yielded a posterior distribution for the detector polarisation angle (PA) that was single-peaked, as shown in the upper-left corner of Figure \ref{fig:simulation} a. To account for the larger error bars observed in real data, we systematically increased the error bars on the simulated modulation curve to match the observational uncertainties. Interestingly, as the error bars grew, the resultant PA posterior developed a shoulder to the most probable peak, closely resembling the observed double-humped distribution. Meanwhile, the posterior of the polarisation fraction (PF) remained single-peaked but with increased standard deviations. These results, thus, suggest that the double-humped distribution of PA obtained from the observed data analysis is a consequence of the larger error bars associated with the observed modulation curve.

\bibliographystyle{aasjournal}



\end{document}